\documentclass[aps,prl,twocolumn]{revtex4}
\usepackage{graphicx,amsmath,subfigure}
\begin{document}
\title{Polarized Neutron $\beta$-Decay: the Proton Asymmetry and Recoil-Order Currents}
\author{S. K. L. Sjue}
\affiliation{Department of Physics, University of Washington, Seattle, WA 98195}\date{\today}

\begin{abstract}
We present an analytic, recoil-order calculation of the proton asymmetry from polarized neutron $\beta$-decay.  The differential decay rate in terms of electron energy and proton direction follows, parametrized in terms of the most general Lorentz-invariant hadron current coupled to a left-handed lepton current. Implications for experimental efforts to measure recoil-order currents are discussed.\end{abstract}
\pacs{}
\maketitle
It is possible to calculate the decay distributions of the proton, electron and antineutrino from a polarized, free neutron due to the weak interaction with great precision within the framework of the standard model.  In the case of neutron decay, electromagnetic effects are relatively small.  Consequently, precision measurements of the decay distributions from polarized neutrons are good candidates in the search for new physics. 
Measurements of the correlations from polarized free neutrons in conjunction with the neutron lifetime, $\tau_n$, have been used to study the overall coupling constant, $G_F$, the ratio of the axial vector to vector couplings, to put limits on possible right-handed currents, and to probe for time reversal invariance-violating effects\cite{pdg:04}.  Given the neutron lifetime, $\tau_n\propto[|V_{ud}|^2G_F^2(1+3\lambda^2)]^{-1}$, it is also possible to extract the quark-mixing matrix element $|V_{ud}|$ from measurements of $\tau_n$ and $\lambda$, where $\lambda$ is the ratio of axial vector to vector coupling in the hadron current.  The value of $|V_{ud}|$ has important implications for the unitarity of the CKM matrix in conjunction with $|V_{us}|$ and $|V_{ub}|$ via the constraint $|V_{ud}|^2+|V_{us}|^2+|V_{ub}|^2=1$.
There are ongoing efforts to improve significantly on these measurements\cite{abBA,UCNA,SerA,aspect,enucor}.

The purpose of this paper is to present an anlytical formula for the proton asymmetry from polarized neutron decay including recoil-order effects. It is useful to expand the neutron's differential decay rate in terms of the electron's maximum energy divided by the neutron mass.  We refer to this small, dimensionless quantity as $R$ for recoil, $R\equiv max(E_e/m_n) \approx .001$.  Both kinematic effects and terms in the interaction current proportional to the momentum transfer contribute at ${\cal O}(R)$.  Taking these effects into account will play a role not only in searching for new physics but in extracting the standard-model form factors from combined measurements.

\begin{figure}
\includegraphics[width=.5\textwidth]{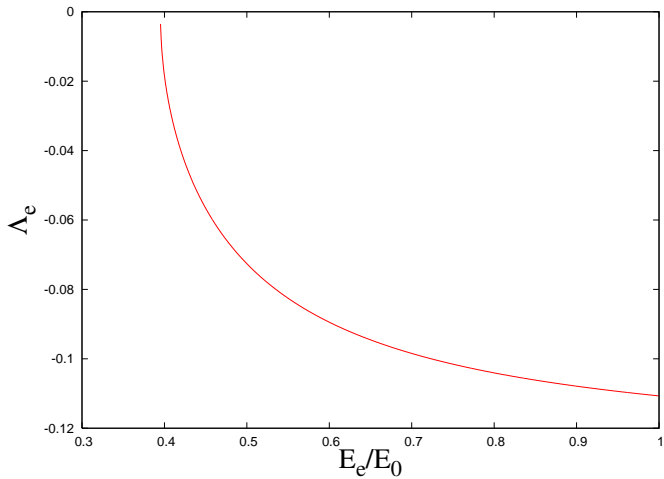}
\caption{The beta asymmetry, with $f_2$ set to its CVC hypothesis value and all other recoil-order hadron couplings set to zero.  The beta asymmetry is dominated by the overall factor $\beta=\frac{p_e}{E_e}$.}
\label{fig:Ab}
\end{figure}

Excellent reviews on the effects of recoil-order corrections in beta decay already exist\cite{Holstein:74,LG:85,Harrington:60} so here we give a brief introduction. A common expression for the decay rate\cite{JTW:57} is 
\begin{widetext}
\begin{equation}
d^5\Gamma=\frac{2|G_F|^2}{(2\pi)^5}E_e |\overrightarrow{p_e}| (E_0-E_e)^2 \Bigg[ 1+
a \frac{\overrightarrow{p_e}}{E_e} \cdot \frac{\overrightarrow{p_\nu}}{E_\nu}+
\overrightarrow{P}\cdot
\left(
A \frac{\overrightarrow{p_e}}{E_e}+
B \frac{\overrightarrow{p_\nu}}{E_\nu}+
D \frac{\overrightarrow{p_e}}{E_e}\times \frac{\overrightarrow{p_\nu}}{E_\nu}
\right)
\Bigg] 
dE_e d\Omega_e d\Omega_\nu,
\end{equation}
\label{melt}
\end{widetext}
in which $E_0$ is the maximum electron energy, $\overrightarrow{p_e}$ and $\overrightarrow{p_\nu}$ are the momenta of the electron and neutrino, and $E_e$ and $E_\nu$ are the energies of the electron and neutrino. $\overrightarrow{P}$ is the neutron's polarization.  As can be seen from the equation $a$ determines the $e-\nu$ correlation, $A$ the beta asymmetry, $B$ the neutrino asymmetry, and $D$ is a T-odd term. The coefficients $a, A, B$, and $D$ depend on the form of the interaction. Within the standard model and ignoring recoil-order effects and radiative corrections, 
\begin{equation}
a=\frac{1-\lambda^2}{1+3\lambda^2}, A=\frac{2\lambda(1-\lambda)}{1+3\lambda^2},\\
\textrm{ and } B=\frac{2\lambda(1+\lambda)}{1+3\lambda^2},\\
\end{equation}
To first order (${\cal O}(R)$), the neutrino exhibits a large asymmetry ($B \approx0.98$) and the electron exhibits a small asymmetry ($A \approx-0.1$, see Figure~\ref{fig:Ab}).  

Because the neutron is a composite object the weak current contains terms in addition to those found for point-like particles and the most general possible (Lorentz invariant) V-A hadron current can be written with six dimensionless constants (form factors), three vector ($f_i$) and three axial vector ($g_i$). Parametrizing these currents in terms of the momentum transfer leads to a matrix element of the form  
\begin{equation}
\textit{M}=\frac{G_F}{\sqrt{2}}\langle p | J^\mu(q^2) | \overrightarrow{n} \rangle \times \overline{e}(p_e)\gamma_\mu(1-\gamma_5)\nu(p_\nu),
\label{eq: matrixelement}
\end{equation}
\begin{widetext}
in which
\begin{equation}
\langle p(p')|J^\mu|n( p , \overrightarrow{s} )\rangle =
\overline{p}(p')\Big[
f_1\gamma^\mu
-i\frac{f_2}{m_n}\sigma^{\mu\nu}q_\nu
+\frac{f_3}{m_n}q^\mu
-g_1\gamma^\mu\gamma_5
+i\frac{g_2}{m_n}\sigma^{\mu\nu}\gamma_5q_\nu
-\frac{g_3}{m_n}\gamma_5q^\mu
\Big]n(p,\overrightarrow{s}).
\label{eq: current}
\end{equation}
\end{widetext}
Here $q^\mu=p^\mu-p'^\mu$ is the momentum transfer, which is equal to the difference between the neutron ($p^\mu$) and proton ($p'^\mu$) momenta.  $m_n$ and $\overrightarrow{s}$ are the neutron's mass and spin.
Because the mass of the neutron is of order $1$ GeV, while the momentum transfer in its decay is $\approx 1$ MeV the recoil-order effects are of order $0.1$\%. All the vector ($f_i$) form factors are related to the isovector electromagnetic form factors of the nucleon via the Conservation of the Vector Current (CVC) hypothesis\cite{FeynmanGellMann:58,GellMann:58}:
\begin{equation}
\begin{array}{ll}
f_1=\hfil&1\hfil\\
f_2=\hfil&\dfrac{\mu_p-\mu_n}{2}\hfil\\
f_3=\hfil&0.\hfil\\
\end{array}
\end{equation}
Both the terms with $f_3$ and $g_2$ are called Second Class Currents (SCC)\cite{Weinberg:58}. Within the standard model and assuming isospin to be an exact symmetry $f_3$ and $g_2$ should be zero, but due to the difference in the quark wave functions within the neutron and proton one expects\cite{do:82,sh:96} $g_2/g_1$ in the range $\approx 0.01-0.05$. Presently the best value of $g_2$ comes from an experiment in the $A=12$ system\cite{Minamisono:02}, which found $2g_2/g_1=-.15\pm.12\pm.05\textrm{(theory)}$.  The pseudoscalar term $g_3$ only results in smaller terms that don't contribute to ${\cal O}(R^2)$.

Measurements of neutron decay have a distinct advantage over experiments with composite nuclei in terms of systematic uncertainties, since one need not account for the effects of the many-body nuclear system. 
In a composite nucleus, the observables used to search for second-class currents include contributions from first-class currents.  In order to disentangle the effects of these two types of couplings, it is necessary to measure both $\beta^+$ and $\beta^-$ decays from mirror nuclei.  It is also necessary to calculate 
and compensate for the two separate nuclear transition matrix elements to the daughter nucleus to use the data from the mirror nuclei.
The neutron is simply three quarks in a bound state.  Precision measurements of the parity-breaking beta and proton asymmetries with respect to the neutron spin could provide better tests of the recoil-order terms within the weak interaction hadron current.  To this end, we present a calcuation of the proton asymmetry.

Much work has been done on recoil-order effects in the weak interaction.  Recoil-order calculations of the lepton asymmetry were performed by Harrington\cite{Harrington:60} for the polarized weak hadron decays of the neutron, $\Sigma^-$, $\Lambda$, and $\Xi$.  A very general treatment within ``effective field theory'' covering various asymmetries and correlations of both composite nuclei and hadrons was published by Holstein\cite{Holstein:74}.  Recently, Gardner and Zhang\cite{GardnerZhang:01} gave results specialized to the neutron for the $\beta$-asymmetry and $e\nu$ correlation.  Gl\"uck and Toth\cite{GluckToth:92} numerically calculated asymmetries, including the recoil asymmetry.  Notably missing from all this work is an analytic calculation of the recoil asymmetry.  We performed an analytic calculation of the recoil asymmetry for completeness, maximum insight into possible systematic errors, and to get access to as many analysis tools as possible for neutron $\beta$-decay.  It is experimentally possible to measure both the electron and the proton from neutron $\beta$-decay.  Several experimental collaborations\cite{Reich:00,Kuznetsov:00,abBA,UCNA,SerA}, are making precision measurements of $A$ and the recoil asymmetry; hopefully calculations of the recoil asymmetry will prove useful in subsequent analyses.

In the process of evaluating the proton asymmetry, it was natural to reevaluate the hadronic matrix element.  We found differences with previous calculations that are listed under \cite{Harrington:60}.  Evaluation of the matrix element in the rest frame of the neutron leads to a general expression of the form
\begin{equation}
\textit{M} = C_1 + 
\overrightarrow{P} \cdot \left(
C_2 \overrightarrow{p_e} + 
C_3 \overrightarrow{p_\nu} + 
C_4  (\overrightarrow{p_e}\times\overrightarrow{p_\nu})
\right),
\end{equation}
in which each $C_i$ is a function of the four-momenta $p_e$, $p_p$, and $p_\nu$.  
We performed recoil-order calculations of $a$ and $A$, obtaining agreement with the results of Gardner and Zhang\cite{GardnerZhang:01}.  Experimentally, current values of these parameters are $\lambda=-1.2695\pm.0029$, $a=-0.103\pm.004$, and $A=-0.1173\pm.0013$\cite{pdg:04}. 

The desired new observable is the decay rate in terms of electron energy and proton angle, or $\frac{d^2\Gamma}{dE_ed(\cos\theta_p)}$.  The easiest way to calculate this is to first integrate over $d^3\overrightarrow{p_\nu}$, then $d(\cos\theta_{ep})$.  In order to obtain the asymmetry term $C_3$ as a function of $\overrightarrow{p_p}$ instead of $\overrightarrow{p_\nu}$, simply substitute $\overrightarrow{p_\nu}=-\overrightarrow{p_e}-\overrightarrow{p_p}$.  With the limits $\cos(\theta_{ep})=\pm1$, conservation of energy and momentum give three limiting equations,
\begin{equation}
\begin{array}{lll}
|\overrightarrow{p_\nu}|=E_\nu=&(m_n-E_e-E_p)=&|\overrightarrow{p_e}|+|\overrightarrow{p_p}|,\hfil \\
&&|\overrightarrow{p_e}|-|\overrightarrow{p_p}|\hfil,\textrm{ and}\\
&&|\overrightarrow{p_p}|-|\overrightarrow{p_e}|.\\
\end{array}
\end{equation}  
The first two provide lower limits of the integral over proton momentum for low and high electron energies, respectively, and the last is an upper limit for all electron energies.  The first of the two lower limits applies when $\overrightarrow{p_e}$ is smaller than $\overrightarrow{p_\nu}$, which is equivalent to $E_e < E_e^c$, where $E_e^c$ is the solution to $\overrightarrow{p_e}=\overrightarrow{p_\nu}$.  The second lower limit applies when $\overrightarrow{p_e}$ is larger than $\overrightarrow{p_\nu}$, or when $E_e > E_e^c$.  These limits reflect the fact that in the neutron's rest frame at very low electron energies, the recoil momentum must oppose the neutrino momentum; similarly at high electron energies, the recoil momentum must oppose the electron's momentum.

It is simplest to express the result in terms of the dimensionless recoil variables.  To this end, we define
\begin{equation}
\begin{array}{lll}
\hfil R\equiv& \dfrac{E_0}{m_n} &=\dfrac{m_n^2+m_e^2-m_p^2}{2m_n^2}\approx.0014,\\
\hfil x\equiv&\dfrac{E_e}{E_0} &=E_e/(Rm_n),\hfil\\
\hfil \epsilon\equiv&\Big(\dfrac{m_e}{m_n}\Big)^2& \approx 3\cdot 10^{-7},\textrm{ and}\\
\hfil \beta\equiv&\dfrac{p_e}{E_e},&\\
\hfil x^c\equiv&\dfrac{E_e^c}{E_0}&=\dfrac{m_n[(m_n-m_p)^2+m_e^2]}{(m_n-m_p)[(m_n-m_p)(m_n+m_p)+m_e^2]}\\
&&\approx0.578\\
\end{array}
\end{equation}
and the limits for the integral over proton momentum become
\begin{equation}
\begin{array}{lll}
             p_p/m_n\equiv&y                                         &\\
             y_-&=\dfrac{R(1-x)}{1-Rx(1+\beta)}-\beta Rx &(x<x^c)\\
             y_-&=\beta Rx- \dfrac{R(1-x)}{1-Rx(1+\beta)}&(x>x^c)\\ 
             y_+&=\beta Rx+ \dfrac{R(1-x)}{1-Rx(1-\beta)}&(\textrm{upper limit  }\forall\textrm{  } x ).\\
\end{array}
\end{equation}

\begin{figure}
\includegraphics[width=.5\textwidth]{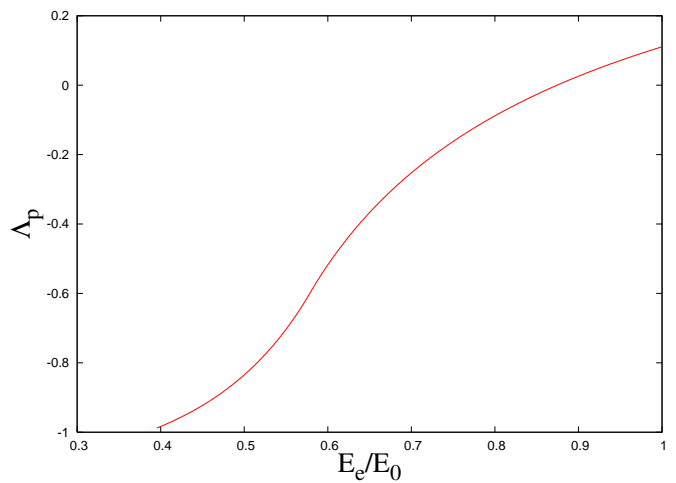}
\caption{The proton asymmetry, with $f_2$ set to its CVC hypothesis value and all other recoil-order hadron couplings set to zero.  $\Lambda_p$ is equal to the observable $2(N_{p+}-N_{p-})/(N_{p+}+N_{p-})$.}
\label{fig:Ap}
\end{figure}

Two integrals are necessary to obtain the proton asymmetry, one for the portion dominated by the neutrino ($E_e$ small) and one for the portion dominated by the electron.  The results for the proton asymmetry follow (see Appendix), with all recoil-order terms included.  See Figure~\ref{fig:Ap} for a plot of the proton asymmetry.  All plots are of the observable 
\begin{eqnarray}
\Lambda=2(N_+-N_-)/(N_+ + N_-),
\end{eqnarray}
where $N_+$ is the number of the given particle emitted in the hemisphere defined by a positive dot product with the direction of the neutron's polarization, and $N_-$ is the number in the opposing hemisphere.  $\Lambda$ is 1 if the given particle is always emitted along the parent's polarization, 0 if the particle is emitted isotropically, and -1 if all emissions oppose the parent's polarization.  Note that the value of the proton asymmetry ranges from $-\Lambda_\nu$ at $E_e=m_e$ to $-\Lambda_e$ at $E_e=E_0$.

\begin{figure}
\includegraphics[width=.5\textwidth]{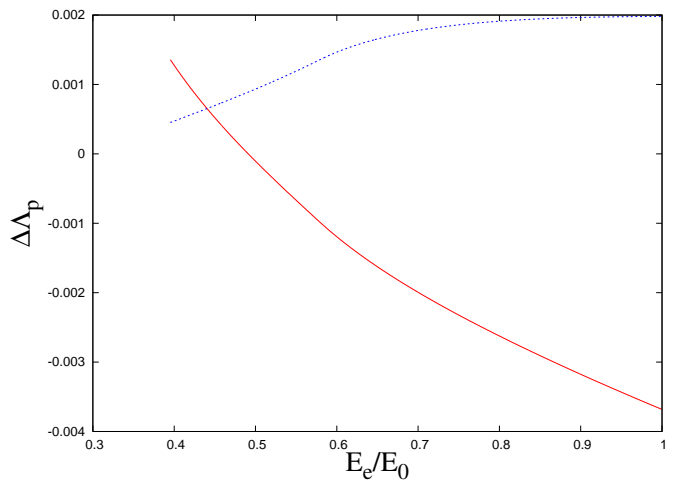}
\caption{Possible changes in the proton asymmetry.  The solid line is the change in $\Lambda_p$ from $f_2$ set to the value predicted by the CVC hypothesis to $f_2=0$.  The dashed line is the change in $\Lambda_p$ from $\lambda$ equal to the world average\cite{pdg:04} to $\lambda$ set to the world average plus its uncertainty, $\lambda+\Delta\lambda$.}
\label{fig:dAp}
\end{figure}

\begin{figure}
\includegraphics[width=.5\textwidth]{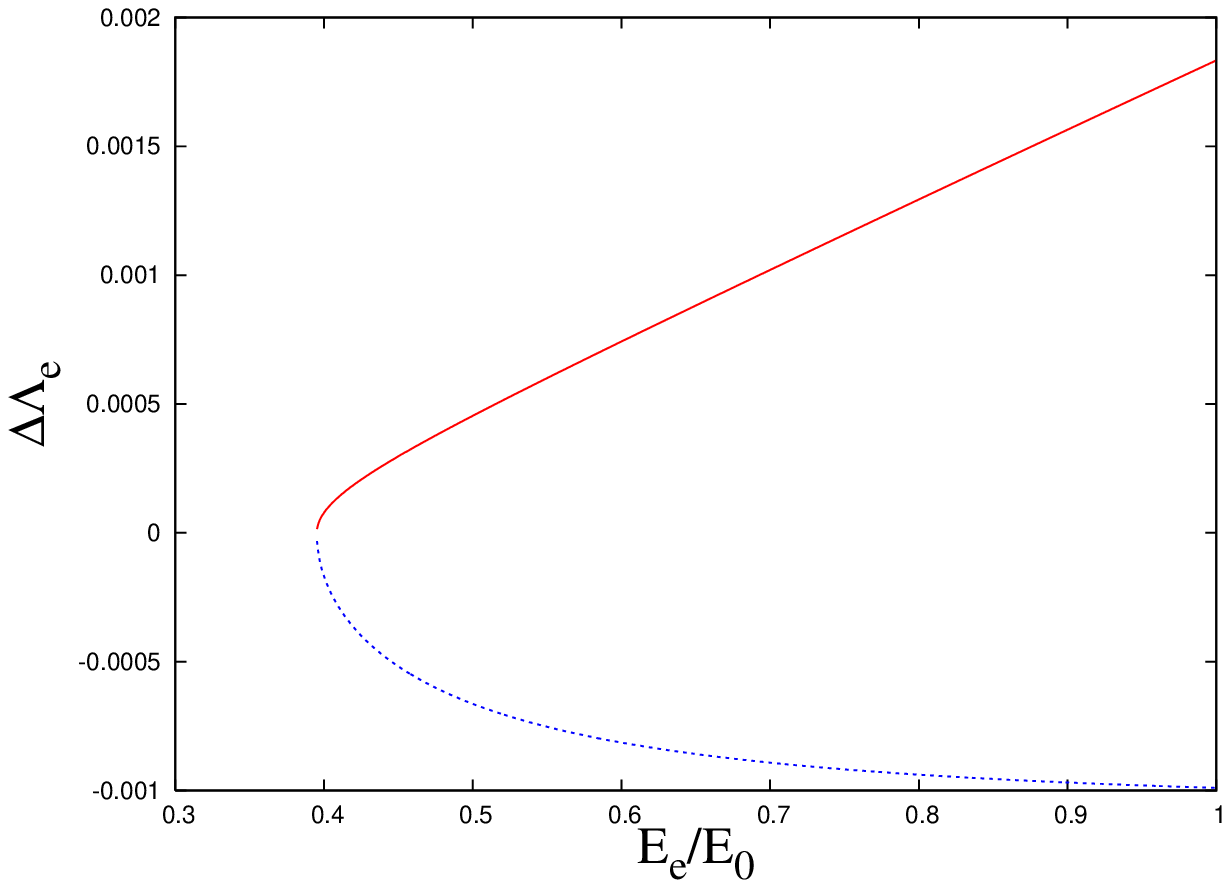}
\caption{The possible changes in the beta asymmetry.  The solid line is the change in $\Lambda_e$ from $f_2$ set to the value predicted by the CVC hypothesis to $f_2=0$.  The dashed line is the change in $\Lambda_e$ from $\lambda$ equal to the world average to $\lambda$ set to the world average plus its uncertainty, $\lambda+\Delta\lambda$.}
\label{fig:dAb}
\end{figure}

The proton asymmetry could be used to measure $f_2$ and check its agreement with the CVC hypothesis.
The absolute magnitude of the $f_2$ contribution to $\Lambda_p$(Figure~\ref{fig:dAp}) is approximately twice as large as the $f_2$ contribution to $\Lambda_e$, the beta asymmetry (Figure~\ref{fig:dAb}).  The overall magnitude of the proton asymmetry is much larger, but the $f_2$ contribution results in a shift of 1.896 keV in the electron energy at which $\Lambda_p$ crosses zero, which could be detected with sufficient precision.  The proton distribution is isotropic at a higher electron energy if $f_2=0$.

SCC effects would be much harder to observe.  Based on the current limit, $g_2$ could only contribute to $\Lambda_p$ at $~5\%$ of the level at which $f_2$ does.  To extract $g_2$ from a measurement of $\Lambda_p$ would require accuracy better than one part in ten thousand.

\begin{figure}
\includegraphics[width=.5\textwidth]{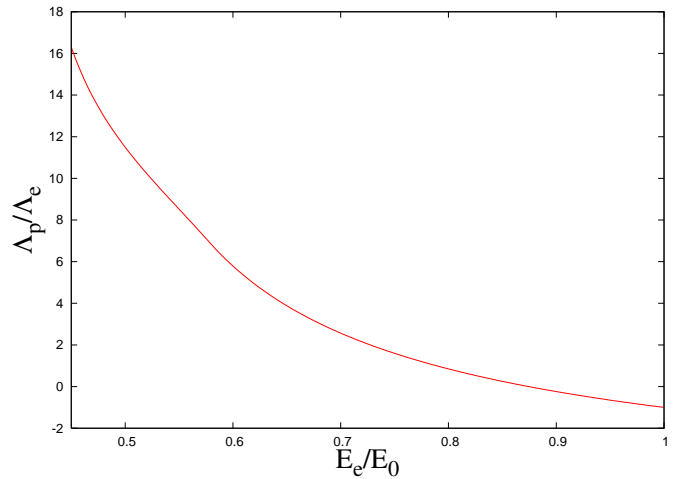}
\caption{The ratio $\Lambda_p/\Lambda_e$, which is independent of the neutron's polarization, with $f_2$ set to its CVC hypothesis value and all other recoil-order hadron couplings set to zero.  The plot excludes the lowest energies because the ratio diverges as $E_e\rightarrow m_e$ and $\Lambda_e\rightarrow 0$.}
\label{fig:ratio}
\end{figure}

\begin{figure}
\includegraphics[width=.5\textwidth]{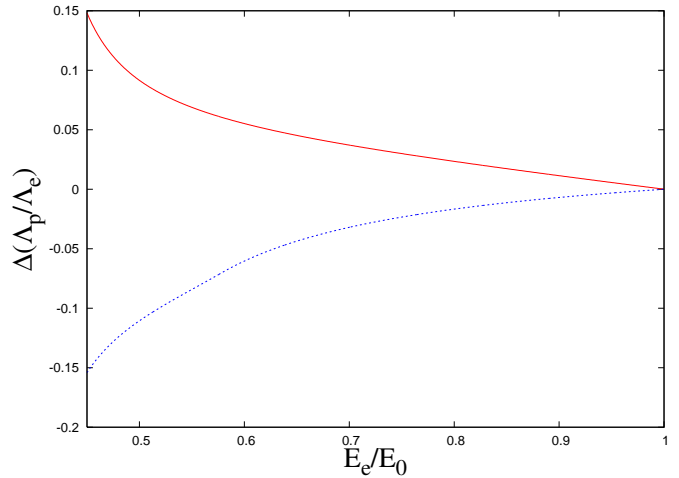}
\caption{Changes in the ratio $\Lambda_p/\Lambda_e$.  The solid line is the change in the ratio from $f_2$ set the value predicted by the CVC hypothesis to $f_2=0$.  The dashed line is the change in the ratio from $\lambda$ equal to the world average to $\lambda$ set to the world average plus its uncertainty, $\lambda+\Delta\lambda$.}
\label{fig:dratio}
\end{figure}

Incomplete knowledge of the polarization of the neutron could be a dominant systematic effect in experiments to measure decay asymmetries~\cite{Kuznetsov:00}, so it is useful to consider a quantity that is independent of the polarization.
The ratio $\Lambda_p/\Lambda_e$ is independent of the neutron's polarization.  
  Figure~\ref{fig:ratio} shows the ratio $\Lambda_p/\Lambda_e$.  $\Lambda_p/\Lambda_e$ also shows sensitivity to the values of $f_2$ and $\lambda$.  Figure~\ref{fig:dratio} shows the change in $\Lambda_p/\Lambda_e$, which is at the 1\% level.  So not only is the ratio of the asymmetries independent of the neutron's polarization, it is also more sensitive to variations in the parameters $\lambda$ and $f_2$ than either $\Lambda_p$ or $\Lambda_e$ alone.

In summary, we presented an analytical expression for the proton asymmetry from polarized neutron decay and used it in conjunction with a similar expression for the beta asymmetry to highlight advantages of a combined measurement.

\section{Acknowledgement}
The author would like to thank Alejandro Garc\'ia and Stephen Ellis for valuable conversations.  This work was performed with support from the Department of Energy under contract DE-F602-97ER41020.

\section{Appendix}

The proton asymmetry follows, omitting a factor of $|f_1|^2$ so that $f_1$ is normalized to 1.  The equations appear as if all form factors are real for the sake of brevity.  To obtain the more general complex expressions, first separate all possible factors of $\lambda^2$ and replace with $|\lambda|^2$.  All remaining expressions involve only two form factors.  Take the real part of the product of one form factor and the complex conjugate of the other, {\em e.g.} $f_2f_3\rightarrow Re(f_2f_3^*)$.  (The only possible exception is a single factor of $\lambda$, which would imply $Re(f_1g_1^*)$.)

\begin{equation}\frac{d^2\Gamma}{dE_ed(\cos\theta_p)}=\frac{2|G_F|^2}{(2\pi)^3} (m_nR)^4 \beta x^2 (1-x)^2[1+ A_p \cos\theta_p]\\
\end{equation}
\vfil

\begin{widetext}
\begin{equation}
\begin{array}{llll}
\hfil A_p =&\hfil -\dfrac{2\lambda}{3(1-x)^2(1 + 3\lambda^2)}\times&[3 \lambda (1-x)^2+3 (1-x)^2+ \beta^2 x ((2-3x)+\lambda(-2+x))]&\\
\\
           &\hfil+ R\dfrac{2}{3(1-x)^2(1 + 3 \lambda^2)^2}\times & \{\lambda[3 (1-x)^3 (\lambda^3-\lambda^2 -\lambda + 1)+\beta^2x(\lambda^3(-5+3x-4x^2-4\beta^2x+ \frac{19}{5}\beta^2x^2)&\\
           &&+\lambda^2(9-11x+4x^2-\frac{11}{5}\beta^2x^2)+\lambda(-3+5x-4x^2+4\beta^2x-\frac{27}{5}\beta^2x^2)&\\
           && +(-1+3x-4x^2+\frac{3}{5}\beta^2 x^2))]&\\
           &&+f_2\lambda[3\lambda^2(1-x)^2(x+2)+3\lambda(1-x)^2(3x-4)+6\lambda(1-x)^3&\\
           &&+\beta^2x(\lambda^2(1-2x)(10x-7)+\lambda(7-8x+2x^2)-6(1-x)^2)&\\
           &&+\beta^4x^2(\lambda^2(-8+\frac{53}{5}x)+\lambda(6-11x)+(2-\frac{4}{5}x))]&\\
           &&+2f_2^2\lambda[\lambda(-3(1-x)^3+\beta^2(1-x)(3-4x+2x^2)+\beta^4(1-x)(x-2))&\\
           &&-3(1-x)^3+\beta^2(1-x)(3-8x+6x^2)+\beta^4x(1-x)(2-3x)]&\\     
           &&+2f_2f_3\lambda^2[-3(1-x)^3+\beta^2(1-x)(3-4x+2x^2)+\beta^4x(1-x)(x-2)]&\\
           &&+f_3\lambda[3\lambda x(1-x)^2+x(1-x)^2+\beta^2x(\lambda (-3+4x-2x^2)+(-3+8x-6x^2))&\\
           &&+\beta^4x^2(\lambda(2-x))+(-2+3x)]      \\
           &&+g_2[2\lambda^3(3(1-x)^3+\beta^2x(1-2x)(1-x)-\beta^4x^2(2-x))&\\
           &&+\lambda^2(3(x-4)(1-x)^2+\beta^2x(1-10x+12x^2))+\beta^4x^2(4-\frac{27}{5}x)&\\
           &&+3\lambda(3(1-x)^2-\beta^2x(2-x))+(3x(1-x)^2+\beta^2x(1-2x)+\frac{1}{5}\beta^4x^3)]\} +{\cal O}(R^2)&\\
           &&(E_e<E_e^c)\\
\\
\end{array}
\end{equation}

\begin{equation}
\begin{array}{llll}
\hfil A_p = &\hfil\dfrac{2\lambda}{3\beta x^2(1+3\lambda^2)}\times&[-\lambda(1-x^2)+(1-3x)(1-x)+3\beta^2x^2(\lambda-1)]&\\
\\
           &\hfil-R\dfrac{2\lambda}{15\beta x^2(1+3\lambda^2)^2}\times& \{(1-x)(2x^2+21x-13)+15\beta^2x^2(1-2x)\\
           &&+\lambda[-(1-x)(3-41x+28x^2)+5\beta^2x(2x+1)(2-3x)-30\beta^4x^3]\\
           &&+\lambda^2[-(1-x)(39-103x+34x^2)+5\beta^2x^2(1-10x)]\\
           &&+\lambda^3[31-28x-7x^2+4x^3+5\beta^2x(-2+3x-10x^2)+30\beta^4x^3]\\
           &&+f_2\lambda[-3(1-x)^2(3x+2)+5\beta^2x(1+4x-8x^2)-15\beta^4x^3]\\ 
           &&+10f_2^2\lambda[(3x-1)(1-x)^2+\lambda(x+1)(1-x)^2\\
           &&+\beta^2(1-x)(1-4x+6x^2-\lambda(1+2x^2))+3\beta^4x^2(1-x)(1-\lambda)]\\
           &&+10f_2f_3\lambda[(1-x)^2((3x-1)+\lambda(x+1)+\beta^2(1-x)((1-4x+6x^2)-\lambda(2x^2+1))\\
           &&+3\beta^4x^2\lambda(1-x)(1-\lambda)]\\
           &&+5xf_3\lambda[-(1-x)(3x-1)+\lambda(x^2-1)+\beta^2((-1+4x-6x^2)+\lambda(1+2x^2))\\
           &&+3\beta^4x^2(1-\lambda)]\\
           &&+g_2[-2(1-x)(3x^2+4x-2)+10\lambda(x^2-1)+5\lambda^2x(x^2-1)-10\lambda^3(x+1)(1-x)^2\\
           &&+10\beta^2x(-x+x^2+3\lambda x+\lambda^2(1-x+9x^2)+\lambda^3(1-x)(2x-1))\\
           &&+15\beta^4x^3\lambda(1-\lambda)]\}+ {\cal O}(R^2)\\
           &&(E_e>E_e^c)\\
\end{array}
\end{equation}
\end{widetext}

\end{document}